\documentclass[pra,onecolumn,notitlepage,twocolumn]{revtex4-1}

\usepackage{amsmath}
\usepackage{amsfonts}
\usepackage{tikz}
\usepackage{graphicx}
\usepackage{subfigure}
\usepackage{pdfsync}
\usepackage{bm}
\usepackage{color}

\newcommand{\ket}[1]{\ensuremath {|\: #1 \: \rangle}}
\newcommand{\bra}[1]{\ensuremath{\langle \: #1 \:|}}
\newcommand{\braket}[2]{\ensuremath{\langle \: #1 \: | \: #2 \: \rangle}}
\newcommand{\ketbra}[2]{\ensuremath{| \: #1 \:\rangle \langle \: #2 \:  |}}

\newcommand{\eref}[1]{(\ref{#1})}
\newcommand{\fref}[1]{figure \ref{#1}}
\newcommand{\Fref}[1]{Figure \ref{#1}}
\newcommand{\sref}[1]{section \ref{#1}}

\newcommand{\llrr}[1]{\ensuremath{\left( #1\right)}}
\newcommand{\llrrq}[1]{\ensuremath{\left[ #1\right]}}

\begin{document}

\title{Noise-assisted quantum transport and computation}


\author{Diego de Falco}
\author{Dario Tamascelli}
\affiliation{Dipartimento di Informatica, Universit\`a degli Studi di Milano\\
Via Comelico, 39/41, 20135 Milano- Italy}
\email{defalco@di.unimi.it,tamascelli@di.unimi.it}



\begin{abstract}
The transmission of an excitation along a spin chain can be hindered by the presence of small fixed imperfections that create  trapping regions where the excitation may get caught (Anderson localization). A certain degree of noise, ensuing from the interaction with a thermal bath, allows to overcome localization (noise-assisted transport). In this paper we investigate the relation between noise-assisted transport and (quantum) computation. In particular we prove that noise does assist classical computation on a quantum computing device but hinders the possibility of creating entanglement.
\end{abstract}


\maketitle

\section{Introduction}
Decoherence and relaxation induced on a quantum system by a fluctuating environment can considerably affect its dynamical behavior. The quantum transport of a particle in a one dimensional tight-binding model, for example, requires coherent hoppings between nearest neighbor locations. The presence of imperfections appearing as a random time-independent potential can hinder the transmission. In fact, the energy mismatches due to such imperfections lead to destructive interference of the wave function \cite{keating06} and Anderson localization \cite{anderson58} appears.
\\However, it has been recently argued that, in certain situations, the presence of an environment can assist the transport of an excitation \cite{aspuru09bis, plenio08, plenio10}. The authors consider exciton transport on small complexes, such as the Fenna-Matthew-Olson (FMO) protein complex and engineered quantum systems (binary trees)  showing that, at room temperature, a limited amount of decoherence (accounting for the interaction between the complexes and the phonons of the surrounding environment) cancels out the localizing effect of small random imperfections along the transmission line. See also \cite{brumfiel12}.
\\[5pt] In this note, we apply similar considerations to the model of universal quantum computer put forward by Feynman in one of the earliest paper on quantum computation \cite{feyn86}. This computer model suits very well the  investigation of the relation between transport and computation, since the transmission of an excitation  does correspond to the execution of a  computation.
\\The  device consists  of an array of two level systems with nearest-neighbor interactions mediated by ancillary qubits. The ancillae play the role of an input/output register. An excitation traveling along the chain plays the role of  a \emph{cursor} administering the applications of unitary transformations (computational primitives) to the register.
In particular, when the cursor reaches the end of the chain, the register is in the output state and the computation is complete. 
\\[5pt]We investigate the effects of irregularities in the on-site energies of the cursor chain and the possibility of assisting the transmission of the cursor along the chain by means of an external potential  and of energy exchanges with the environment. We show that the cursor does indeed travel beyond the localization length and reaches the far end of the chain. We discuss the conditions on the system-environment interaction and on the kind of information processed that allows for an extension of noise-assisted transport to noise-assisted computation.
\\[5pt] The paper is organized as follows: we first discuss the detrimental effect of random static noise on transmission (section \ref{sec:model}) and present an instance of noise-assisted transport by means of energy exchanges (section \ref{sec:noise-assisted}). We then discuss the effects of noise on the computational capabilities of the Feynamn quantum computer (section \ref{sec:computation}). The last section is reserved to discussion and outlook.

\section{Localization on a linear lattice} \label{sec:model}
\begin{figure*}[h]
$\centering
\begin{array}{c}
\subfigure[]{\includegraphics[width=0.9\columnwidth]{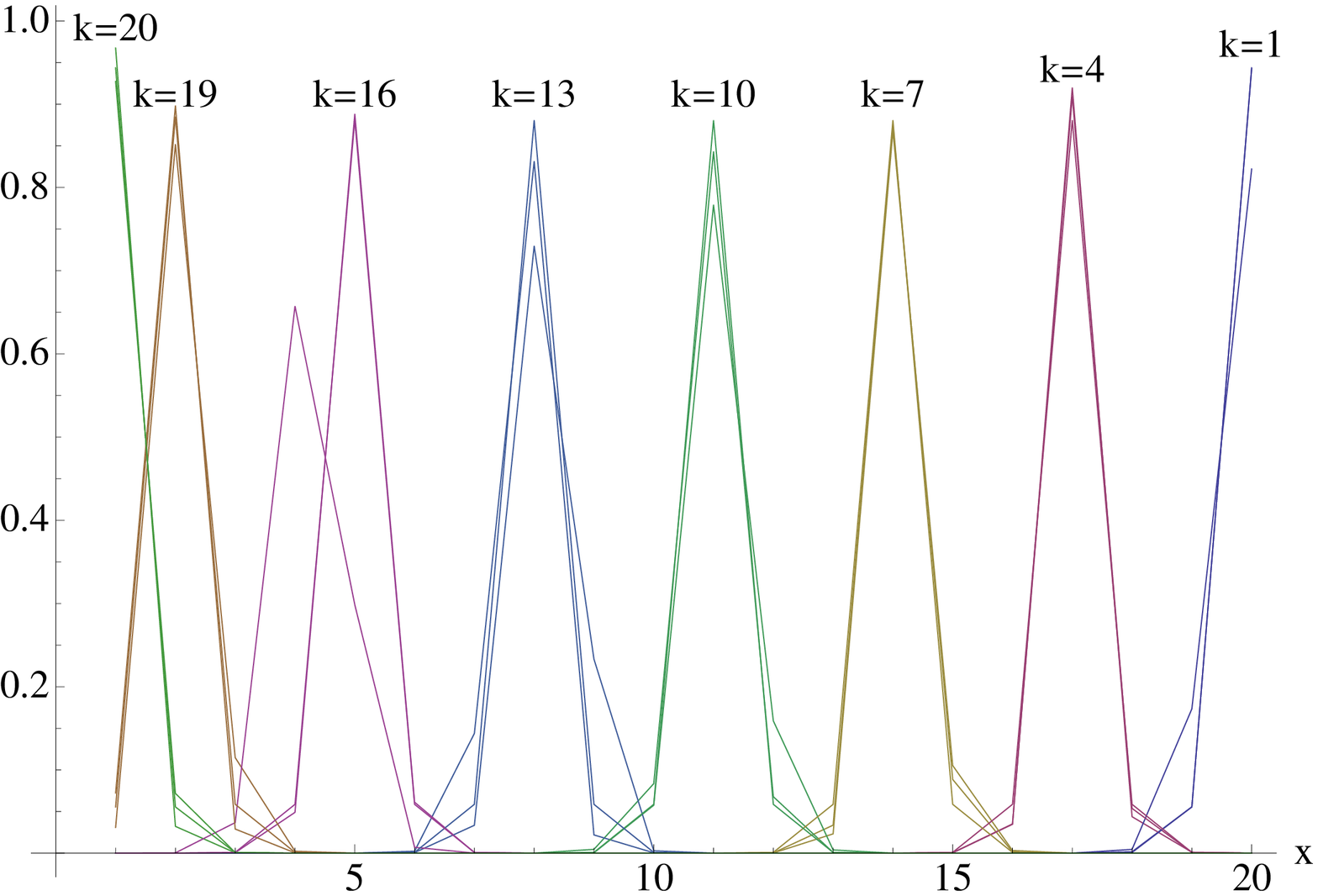}} \hspace{0.1\columnwidth}
\subfigure[]{\includegraphics[width=0.9\columnwidth]{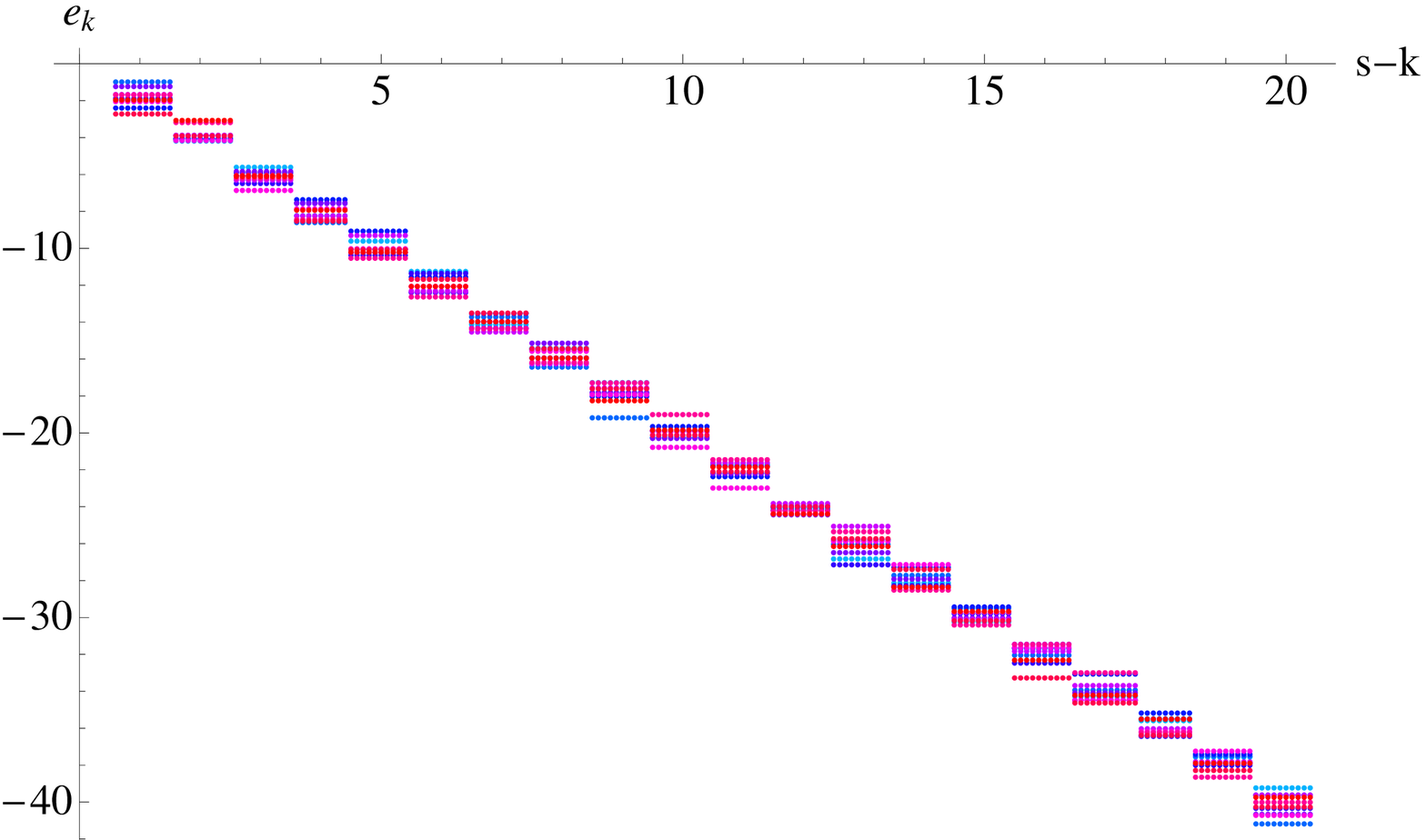}}
\end{array}$
\caption{
$s=20$; $g=2$; . 
(a) The probability distributions of the observable $Q$ in some eigenstates for two realizations of $H_{R+L}$  with $\sigma=0.5$ and  one with $\sigma=0$ (no imperfections). (b) The eigenvalues of $H_{R+L}$ for a sample of 100 realization of the random imperfections for $\sigma=0.5$.
\label{fig:blochStates}}
\end{figure*}

Consider an excitation moving along a one dimensional regular lattice, e.g. an electron moving along a periodic potential, in the tight-binding approximation. Calling \ket{x} the state localized at site $x$,  the Hamiltonian for a lattice of $s$ sites is:
\begin{align}\label{eq:H0}
 H_0 = -\frac{1}{2}\sum_{x=1}^{s-1} \ketbra{x+1}{x}+\ketbra{x}{x+1}.
\end{align}
The evolution \ket{\psi_t} of an initial condition \ket{\psi_0} is:
\[
\ket{\psi_t} = \sum_{k=1}^s \exp\llrr{-i t e_k}  \ket{e_k}\braket{e_k}{\psi_0}
\]
where:
\begin{align}
 &e_k=-\cos \left (\frac{k \pi}{s+1} \right), \ k=1,2,...,s \label{eq:freeEval}\\
 &\ket{e_k} = \sqrt{\frac{2}{s+1}}\sum_{x=1}^s \sin\left (\frac{k \pi x}{s+1} \right) \ket{x}.
 \end{align}
We are interested in the transport capability of the chain, that is in the probability that an excitation initially located at the left hand of the chain (\ket{\psi_0}=\ket{1}) reaches the rightmost site $s$, namely in the probability
\[
P_{Q=s}(t)=\left | \braket{s}{\psi_t}\right|^2,
\]
where $Q$ is the position operator 
\[
Q=\sum_{x=1}^s x \ketbra{x}{x}.
\]
\begin{figure*}[h]
$\begin{array}{c}
\subfigure[]{\includegraphics[width=0.9\columnwidth]{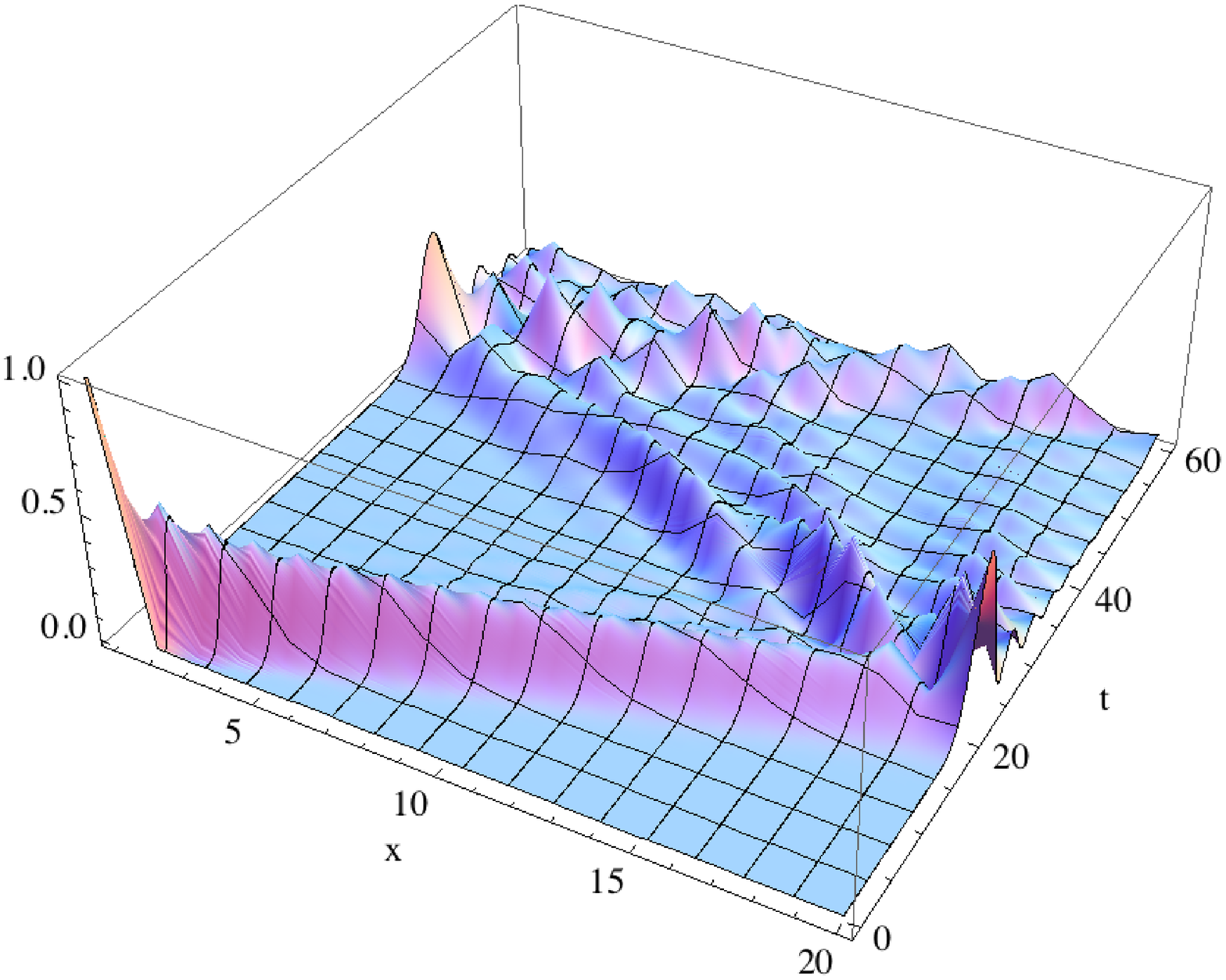}} \hspace{0.1\columnwidth}
\subfigure[]{\includegraphics[width=0.9\columnwidth]{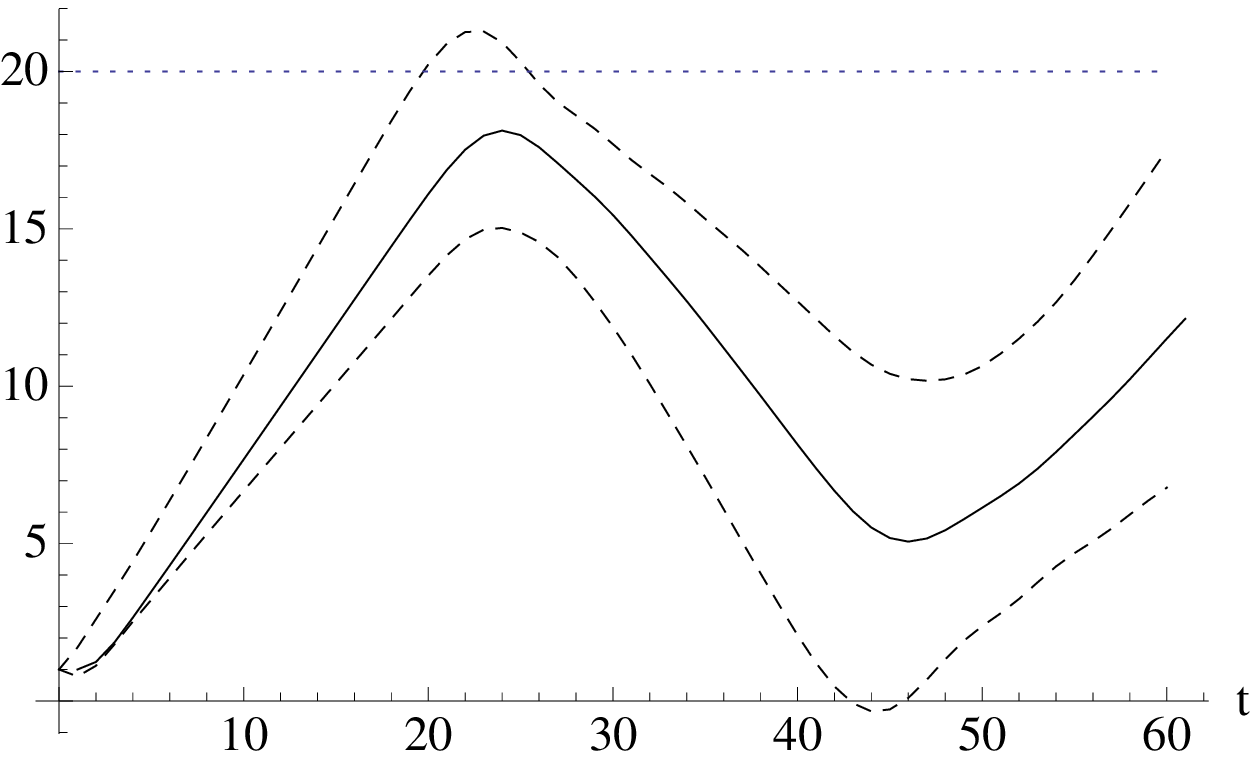}} 
\end{array}$
\caption{$s=20$; (a) The initial condition $\ket{\psi_0}=\ket{1}$ evolving under the Hamiltonian $H_0$. (b) The expectation $E(Q)=\bra{\psi_t}Q \ket{\psi_t}$ as a function of time (solid line) together with $E(Q)\pm \sqrt{var(Q)}$ (dashed lines). \label{fig:figFree}}
\end{figure*} 
\Fref{fig:figFree} shows an example of the motion of the excitation along the lattice: the particle moves ballistically along the chain until it bounces against the boundaries. As time goes on, the wave packet spreads over the line. The probability of finding the excitation at the end of the chain is highest when the first reflection occurs (around time $t=s$) and, at that time, is of order $O(s^{-2/3})$ \cite{apolloni02}.
\\[5pt]We now turn our attention to the effects  of imperfections on transmission. We model imperfections of the lattice by a random time-independent potential. In the position representation we adopt, the random potential appears in the main diagonal of the Hamiltonian:
\begin{align}
 H_R = H_0 + V_R = H_0+\sum_{x=1}^s \epsilon_x \ketbra{x}{x}.
\end{align}
The coefficients $\epsilon_x, \ x=1,2,...,s$ are independent realizations of a random variable $R$.
\\The fact that the presence of imperfections can localize the excitation/cursor (Anderson localization) is well known \cite{anderson73, mott61,lloyd69,thouless72}. The phenomenon can be understood in different ways. One is in terms of destruction of the coherent hopping between nearest neighbors, that leads to destructive interference of the wave function. Another interpretation, that better suits our discussion, is given in terms of the eigenstates of the Hamiltonian $H_R$: the presence of small imperfections on the chain results in a localization of the eigenstates of $H_R$. In our particular instance, since the initial condition has a non-negligible projection on the eigenstates localized at the left of the chain, and the eigenstates do not change shape (but only phase) in time, the excitation remains localized around its starting position.
\\The typical size of the localized states (localization length) depends on the distribution of the noise. Here we consider the random variable $R$ to have  a Gaussian distribution with mean zero and variance $\sigma^2$. With this choice, the localization length is \cite{malyshev03} 
\[
\ell(\sigma) = \llrr{\frac{2\pi^2}{\sigma}}^{2/3}.
\]
The idea of adding a linear potential $V_L= -g \sum_{x=1}^s x \ketbra{x}{x}$, with $g>0$, to reduce the effect of the irregularities in the chain comes from a classical  prejudice: classically, one expects this  potential to pull the excitation rightward; moreover, with a suitable choice of the magnitude of $g$, the defects (small random potential terms) would become negligible with respect to the linear potential.
\\It is immediate to see that this idea/prejudice is very naive and actually quite misleading: the eigenstates of the Hamiltonian
\begin{align}
H_{R+L} & = H_0 + V_R +V_L = \\
&=H_0 +\sum_{x=1}^s(\epsilon_x -g x) \ketbra{x}{x}, \nonumber
\end{align}
with $g>0$ are still localized. Localization manifests itself, in this case, in the form of Bloch oscillations, a genuinely quantum effect: the momentum of the quasi-particle moving on the lattice changes linearly with time, $p(t)=p(0)-gt$, until it reaches the boundary of the Brillouin zone, where it is Bragg reflected \cite{bloch29}. The localization length is determined by the intensity of the force through \cite{hartman04}:
\[
\ell(g)=\frac{\Delta_{H_R}}{g}
\]
where $\Delta_{H_R}$ is the bandwidth of the Hamiltonian $H_R$.
\\As an example, figure \ref{fig:blochStates} shows distribution of the observable $Q$ for the eigenstates of $H_{R+L}$, for $s=20$, $\sigma=0.5$ and $g=2$: the eigenstates are localized. If the particle is initially localized at site $1$ it has a non-vanishing projection only on the eigenstates localized in the same region, and will therefore remain localized there. 
\\The eigenvalues of the same Hamiltonian for a sample of 100 realization of the random imperfections are shown in \fref{fig:blochStates}(b).  
\\In the presence of imperfection, therefore, the transmission of a particle along a linear lattice via Hamiltonian evolution is suppressed and the idea of pulling the particle rightward by means of an external constant force, mimicking the effect of a battery, does not work.

\section{Noise assisted transport} \label{sec:noise-assisted}
The onset of interaction with a thermal environment might be expected to further reduce the transmission capabilities,  of the system. But this is not necessarily the case. Recent work showed that some interaction with the environment can assist transport on short chains \cite{aspuru09bis} and  quantum networks \cite{plenio08,plenio10}.
\\In this section we investigate the effects of dissipation on the transport capabilities of the linear lattice described in \sref{sec:model}.
\\[5pt]The linear chain is now weakly coupled to a thermal bath. The full Hamiltonian (system+environment) is:
\begin{align}
 H_{TOT}=H_S+H_E+H_I,
\end{align}
where $H_S$ is the Hamiltonian of the chain (either $H_0$ or $H_R$ or $H_{R+L}$), $H_E$ is the Hamiltonian of the environment and $H_I$ describes the system/environment interaction. Following \cite{tannoudji92}, we suppose $H_I= -\mathcal{E} \otimes \mathcal{S}$, where $\mathcal{E}$ is an observable of the environment and $\mathcal{S}$ is an observable of the system.
\\In order to know the specific form of $\mathcal{E}$ and $\mathcal{S}$ we should know the precise implementation of the array and the nature of the environmental degrees of freedom. Since this is not the case, we limit ourselves to formulating the hypotheses under which we derive the results of this and the following section.
\begin{itemize}
\item The environment must act as a reservoir: the effects on it of the interaction with the system are negligible. Moreover, we assume that the reservoir is in a thermal stationary state, that is $\rho_E\propto e^{-\beta H_E}$ at some inverse temperature $\beta$. 
\item The system exchanges energy with the environment, not particles. In other words, the number of excitation in the chain remains a constant of the motion even when it interacts with the reservoir. If the operator $\mathcal{S}$ is of the form:
\begin{align} \label{eq:obS}
\mathcal{S} = \sum_{x=1}^s f(x) \ketbra{x}{x},
\end{align}
for example, this requirement is satisfied. We point out that this choice is not unique but is, by far, the simplest, since it assumes that the bath interacts locally with the chain sites. In the following section this requirement will have significant implications on the computational capabilities of the system as well.%
\item The environment can possess a continuous spectrum, but due to the weak coupling the system couples only with the resonant levels. Moreover we require the density of states of the bath to be constant over the resonant energy levels.
\item The evolution of a given initial state $\rho(0)$ of the system is determined by  the Lindblad equation:
\begin{align} \label{eq:Lindblad}
 \frac{d \rho(t)}{dt} = i \llrrq{\rho(t),H_S} + \mathcal{D}(\rho(t)),
\end{align}
where the dissipator $\mathcal{D}$ has the form:
\begin{align} \label{eq:dissipator}
\mathcal{D}(\rho) &= \zeta \sum_{m=1}^s \sum_{n=1}^s \gamma(m,n) \llrr{L(m,n} \rho L(m,n)^{\dagger} \\
&- \frac{1}{2} L(m,n)^\dagger L(m,n) \rho -\frac{1}{2} \rho L(m,n)\dagger L(m,n).\nonumber
\end{align}
Here $\zeta$ is the (positive) system-environment coupling constant. The generators $L(m,n)$ are defined in terms of the normalized eigenstates of $H_S$ (satisfying: $H_S \ket{n} = e_n \ket{n}; \ e_1<e_2<...<e_s;\ \braket{n}{m}=\delta_{n,m}$):
\begin{align}
 &L(n,n)=0\\
 &L(m,n) = \ketbra{m}{n},\; \mbox{for} \ m \neq n. \nonumber
\end{align}
Each coefficient $\gamma(m,n)$ in \eref{eq:dissipator}, for $m \neq n$, has the meaning of transition probability per unit time from \ket{n} to \ket{m}.
We set, in what follows, for $n \neq m$, $\omega(m,n) = \omega(n,m) = |e_n - e_m|$.
\\We require, following \cite{gebauer04}, that transitions $n \to m$, with $n>m$, in which the system absorbs energy from the reservoir, take place at the  absorption rate:
\begin{align} \label{eq:absorption}
 &\gamma(m,n) = \eta \llrr{\omega(m,n)} \frac{1}{e^{(\beta \ \omega(m,n))}-1}.
\end{align}
Transitions $n \to m$, with $m<n$, in which the reservoir drains energy from the system, take place at a rate
\begin{align}\label{eq:emission}
 &\gamma(m,n) = \eta \llrr{\omega(m,n)} \llrr{\frac{1}{e^{\beta \omega(n,m)}-1}+1},
\end{align}
taking into account both stimulated emission
 and spontaneous emission, via the additional term +1 appearing in \eref{eq:emission}.
\item We consider only the case of an underdamped motion in which simultaneous emission of two or more phonons is forbidden; namely we assume that 
\[
\eta(\omega(m,n)) = \delta_{m,n+1}+\delta_{m,n-1},
\]
but for irrelevant constants that we absorb in the system/environment coupling constant $\zeta$.
\end{itemize}
%
Under these conditions, the equations for the elements of the system density matrix $\rho(t)$, in the energy representation, are the following:
\begin{widetext}
\begin{align}
 &\frac{d}{dt} \rho_{m,n}(t)=\llrr{-i (e(m) - 
    e(n)) - \zeta \sum_{j=1}^s \frac{\gamma(j, m) + \gamma(j, n)}{2}}\rho_{m,n}(t), \; m \neq n \label{eq:off}\\
 &\frac{d}{dt} \rho_{m,m}(t)=  \zeta \sum_{c=1}^s\rho_{c,c}(t) \gamma(m,c) - \rho_{m,m}(t) \gamma(c,m) \label{eq:trans}
 \end{align}
 \end{widetext}
where $e(k), \ k=1,\ldots,s$ are the eigenvalues of the Hamiltonian $H_S$ \cite{tannoudji92}.
\\[5pt]If the Hamiltonian of the system is the free Hamiltonian $H_0$, the transmission capabilities of the computing device is dramatically reduced by the presence of the bath, as expected: the system relaxes toward a thermal state that, the lower the temperature $1/\beta$, the more is concentrated around the middle of the chain.
\\The introduction of disorder is obviously not expected to do any better: the eigenstates of the Hamiltonian $H_R$ are localized around the peaks of the random potential.
\\The situation is radically different when a linear (externally tunable) potential is added to the random one: Bloch localization makes the eigenstate of the Hamiltonian $H_{R+L}$ localized in space (see \fref{fig:blochStates}(a)). The interaction with the bath determines a ``stroll'' over the energy eigenstates which, because of the structure of the bath, favors, at low temperatures, the propagation toward states of lower energy. The localized nature of the eigenstates of $H_{R+L}$, due, we remind, to the presence of the external potential $V_L$, makes the random walk on the energies quasi-equivalent to a random walk on the sites of the chain: losing energy amounts to moving to the right.
\begin{figure*}[h!]
\centering
$\begin{array}{c}
\subfigure[]{\includegraphics[width=0.9\columnwidth]{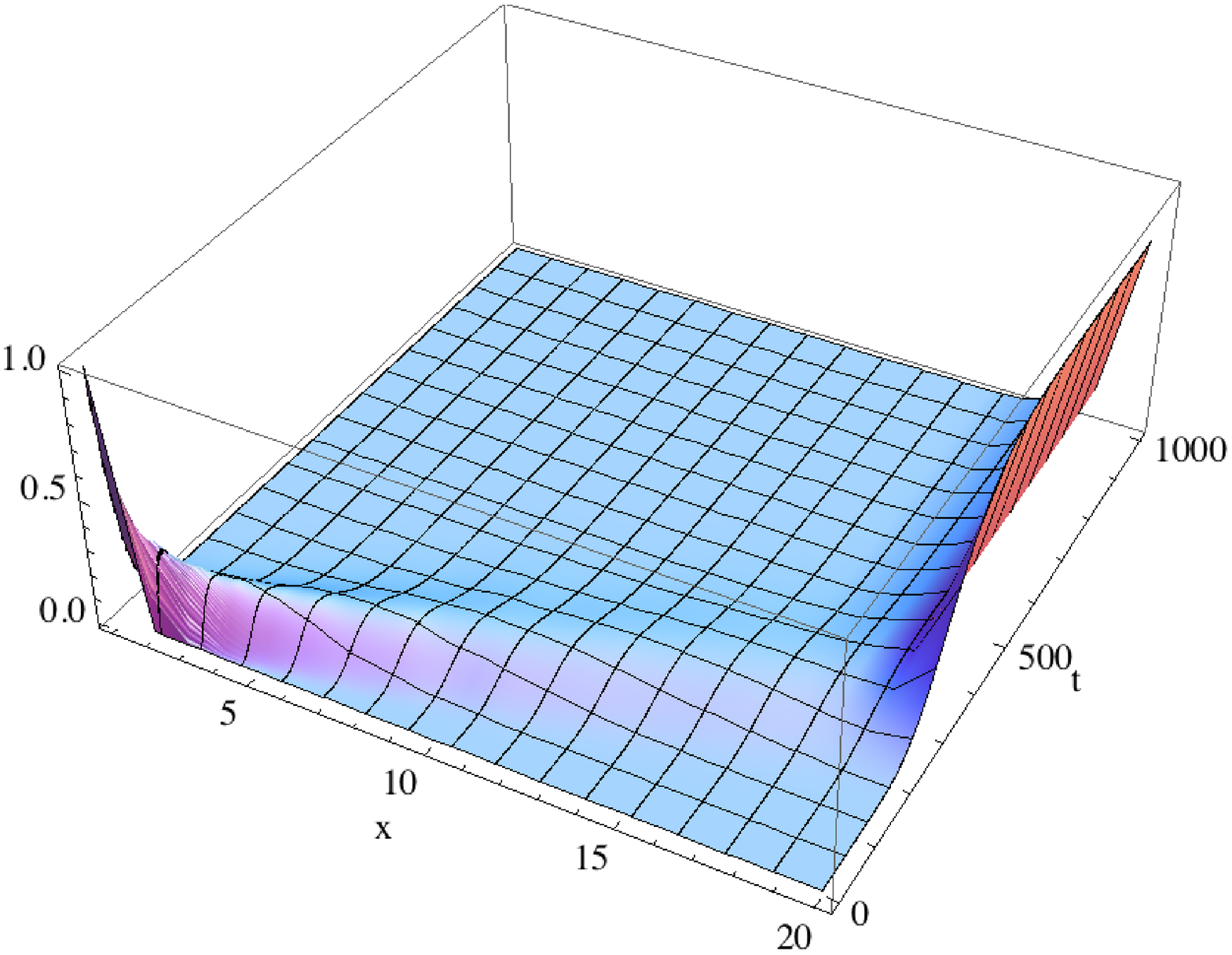}} \hspace{0.1\columnwidth}
\subfigure[]{\includegraphics[width=0.9\columnwidth]{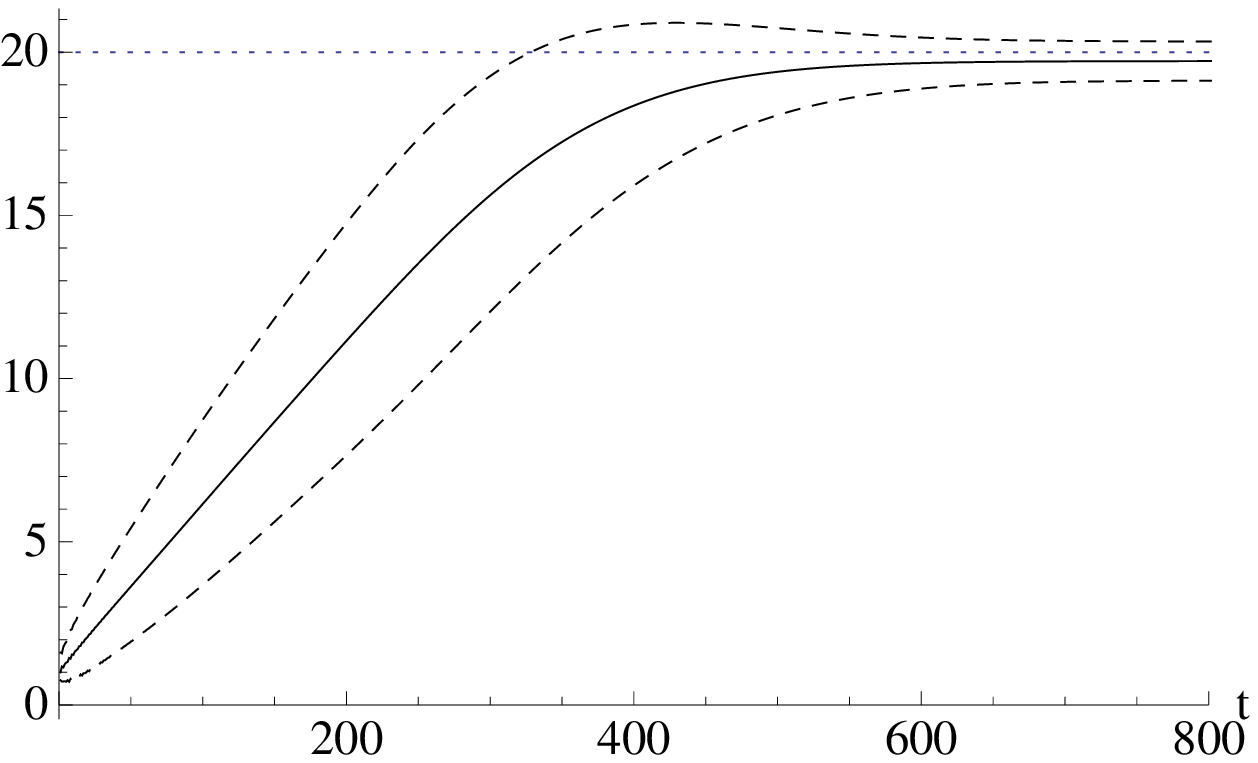}}
\end{array}$
\caption{$s=20$; $\sigma=0.5$. (a) The initial condition $\rho_0=\ketbra{1}{1}$ evolving under $H_{R+L}$, with $g=2$  and (weakly) interacting ($\zeta=0.05$) with a reservoir at inverse temperature $\beta = 1$. (b) The expectation $E(Q)=Tr\llrr{Q \rho(t)}$ as a function of time (solid line) together with $E(Q)\pm \sqrt{var(Q)}$ (dashed lines).\label{fig:figDiss}}.
\end{figure*} 
\\The speed of the walk on the energy landscape induced by the dissipation depends on the energy gaps between the eigenstates of $H_{R+L}$ and on the coupling constant $\zeta$. This dependence can be analytically accessed by inspection of equations \ref{eq:off} and \ref{eq:trans}.
\\The diagonal elements describe a random walk on the energy landscape of the system. The presence of a sufficiently strong external potential  ($g>1$) makes the energy gaps $\left|e(k)-e(k\pm 1) \right |$ almost equal for $k=1,\ldots,s$ (see, for example \fref{fig:blochStates}(b)); the emission (and absorption) rates $\gamma_E(x)=\gamma(x,x+1)$ (and  $\gamma_A(x)=\gamma(x+1,x)$) are then independent of $x$ ($\gamma_E(x)\approx \gamma_E, \gamma_A(x)\approx\gamma_A$). For $\beta$ high enough, $\gamma_A$ becomes negligible and $\gamma_E\approx 1$. Therefore the random walk on energies described by equation \ref{eq:trans} proceeds toward states of lower energy at a rate $\zeta\ \gamma_E \approx \zeta$.
\\The coherences (off-diagonal terms) are described by autonomous equations \eref{eq:off} and are exponentially depressed in time (decoherence).
\\[5pt]Frames (a) and (b) of \fref{fig:figDiss} show the evolution of the initial condition $\rho_0 = \ketbra{1}{1}$ under the Lindbladian \eref{eq:Lindblad} for $\beta=1$, $g=2$ and a realization of the random imperfections extracted from a zero-mean Gaussian with standard deviation $\sigma=0.5$.
\\[5pt]The transport process determined by the Hamiltonian dynamics (see \fref{fig:figFree}) is qualitatively different from the one depicted in \fref{fig:figDiss}.
\\Under ``perfect'' Hamiltonian evolution the particle bounces back and forth. It is thus necessary to determine the optimal time at which to measure the position of the excitation, that is the time at which the probability of finding the excitation at the end of the chain is maximal and this probability decreases with $s$ as $O(s^{-2/3})$. Moreover, the presence of disorder suppresses transmission.
\\In the presence of a bath and an external static field, the transmission of the particle to the far end of the chain takes a time that is one order of magnitude larger than under Hamiltonian evolution in the absence of imperfections (compare the time scales figures \ref{fig:figFree} and \ref{fig:figDiss}). This is due to the weak system-bath coupling ($\zeta=0.05$). But once the excitation has arrived at the end of the chain it remains there, and the motion is stable with respect to random fluctuations of the on-site energies. 
\\The coupling with a cold reservoir, in the presence of a uniform force field that completely offsets a weak random potential, assists the transport (however slow and incoherent) of an excitation along a line.

\section{Noise-assisted (quantum) computation} \label{sec:computation}
\begin{figure*}[h!]
\centering
$\begin{array}{c}
\includegraphics[width=1.\columnwidth]{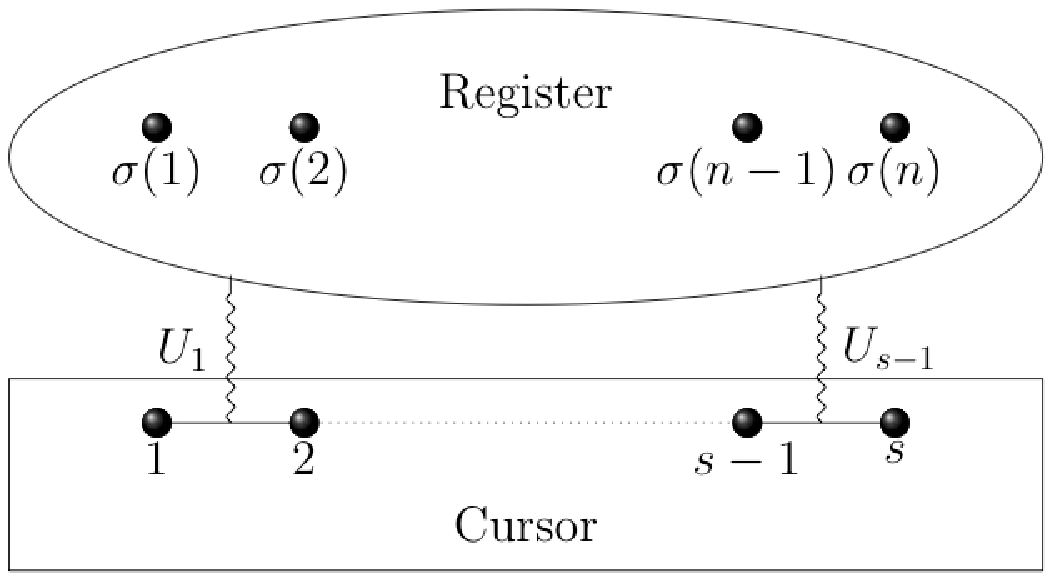}
\end{array}$
\caption{The Feynman machine: the $s$-sites cursor interacts with the register as described in \eref{eq:feyn}.\label{fig:system}}
\end{figure*}

\begin{figure*}[h!]
\centering
$\begin{array}{c}
\includegraphics[width=1.6\columnwidth]{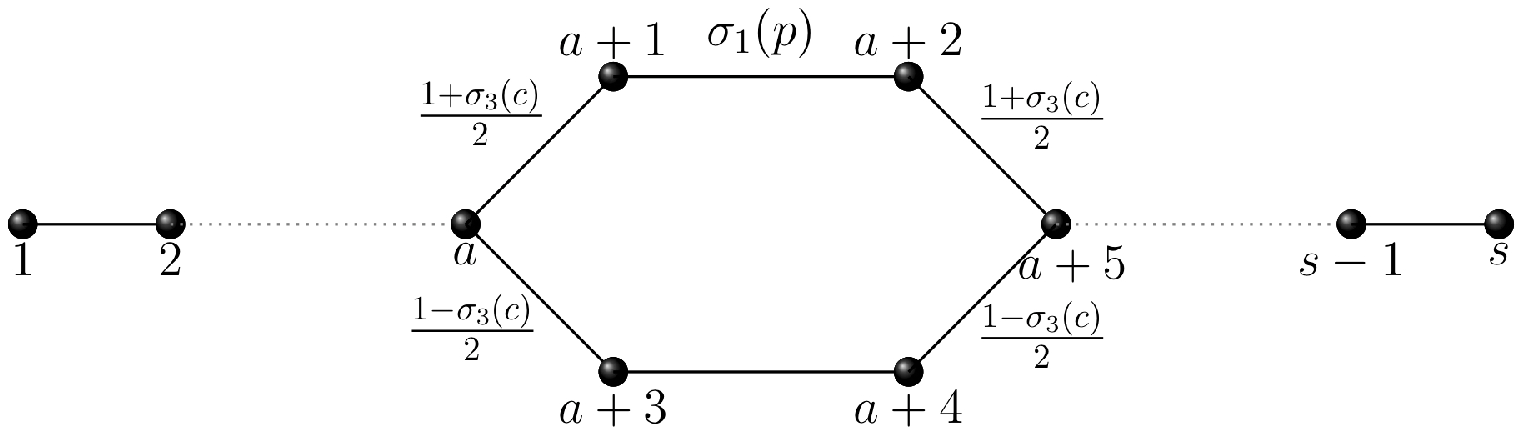}
\end{array}$
\caption{The CNOT described by equation \ref{eq:CNOTtel}. The state of the controlling qubit $\sigma(c)$ determines which branch of the circuit is visited by the moving particle. This determines the transformation applied to the controlled qubit $\sigma(p)$:  $\sigma_1(p)=NOT(p)$ along the upper branch, the identity along the lower branch. }\label{fig:CNOT}
\end{figure*}

\begin{figure*}[h]
\centering
\includegraphics[width=1.3\columnwidth]{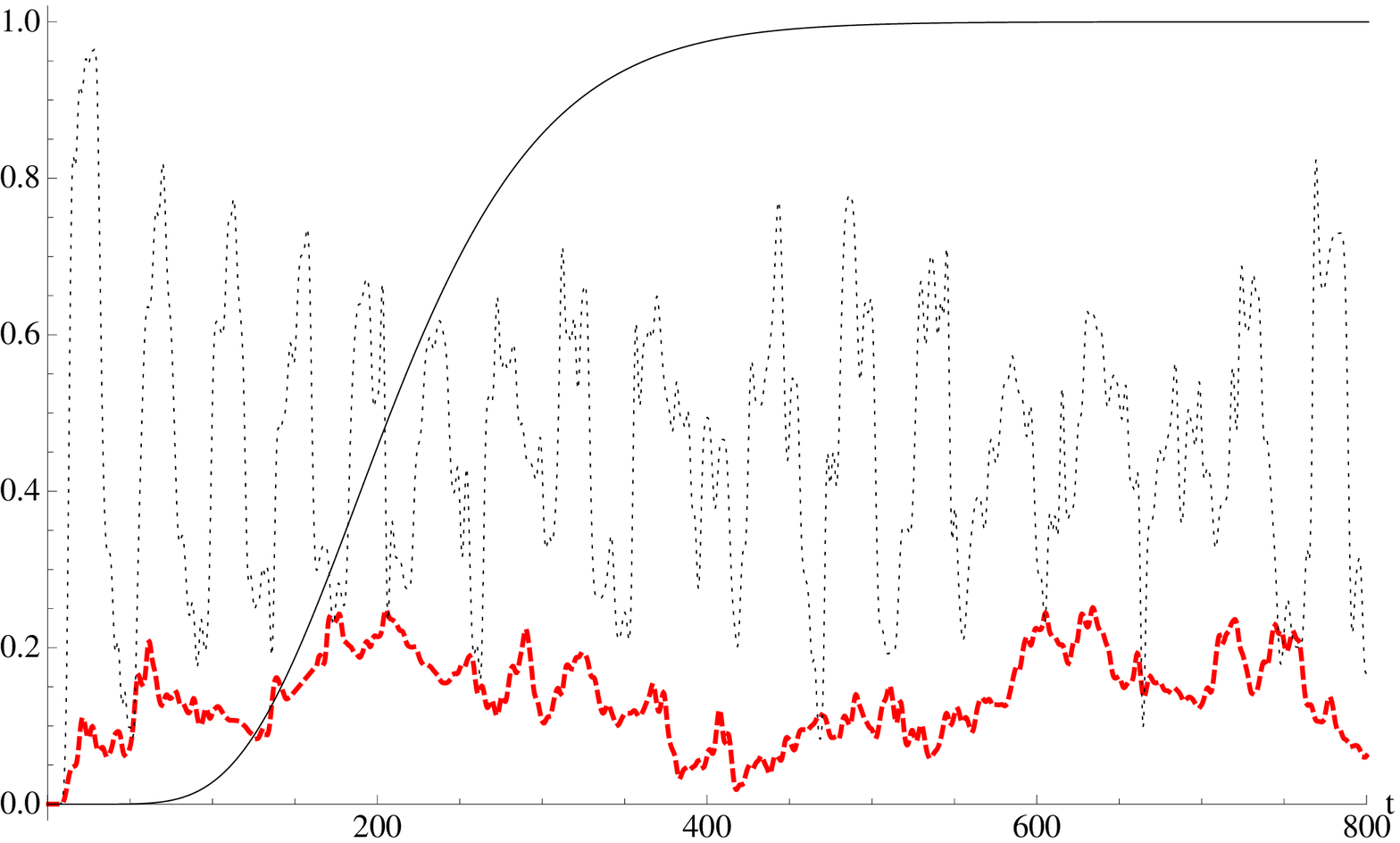}
\caption{$s=22$: with this choice, each computational path consists of $s-2=20$ sites, thus allowing for comparison with  figures \ref{fig:figFree} and \ref{fig:figDiss}; $a=9$. The probability of reaching, starting from site 1, the region $b,\ldots,s$ under three different evolutions. Dotted line:  Hamiltonian evolution determined by $H_F$; dashed line: Hamiltonian evolution under $H_F + V_R$, for a realization of the random potential $V_R$  with $\sigma=0.5$. Solid line: the dissipative evolution determined by the Hamiltonian $H_F+V_L'+V_R$ for the same realization of $V_R$, $g=2$, coupling $\zeta=0.05$ and inverse temperature $\beta=1$.
\label{fig:figNAC}}
\end{figure*}

\begin{figure*}[h]
\centering
\includegraphics[width=1.3\columnwidth]{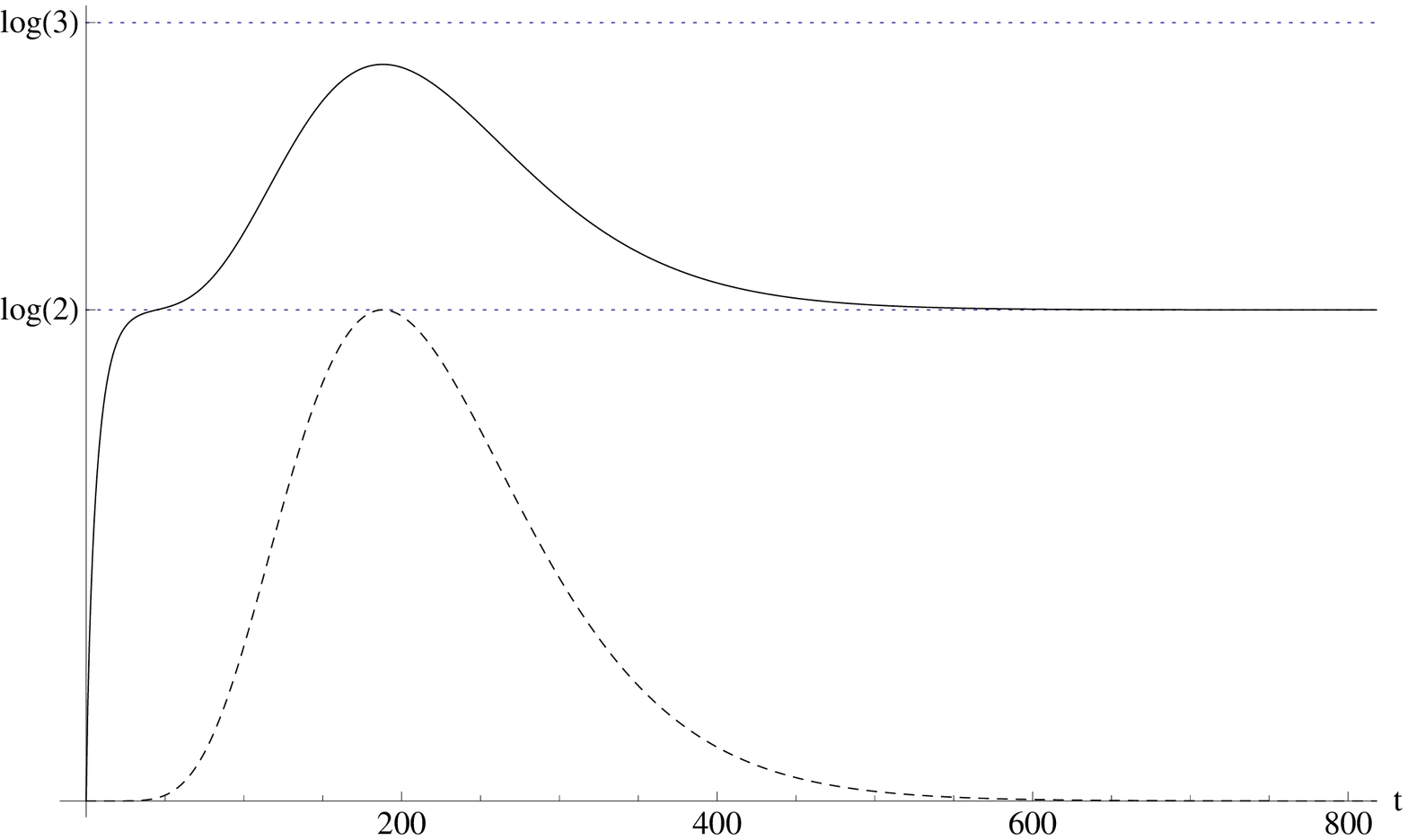}
\caption{$s=22; \ a=9; \ \zeta=0.05; \ \beta=1$. The von Neumann entropy of the register for $\rho^U(t)$ (dashed line) and  $\rho^{U+D}(t)$ (solid line). The von Neumann entropy for completely mixed of 2- and 3-state systems are represented as horizontal lines for comparison purposes.  \label{fig:figFidelityEntropy}}
\end{figure*}

In this section we discuss the relation between quantum transport and quantum computation. The quantum computer model we use in our investigation is the one proposed by Feynman \cite{feyn86}.
\\[5pt]Feynman's quantum computer consists of two logically separated parts; one part, the \emph{clock}, is an excitation moving along a lattice. The second part, the input/output register, is a collection of additional degrees of freedom, say $n$ spin 1/2 particles $\sigma(j)=(\sigma_1(j),\sigma_2(j),\sigma_3(j)),\ j=1,2,\ldots,n$.
\\The system is governed by the \emph{time-independent} Hamiltonian:
\begin{align} \label{eq:feyn}
 H_F = -\frac{1}{2}\sum_{x=1}^{s-1} \ketbra{x+1}{x} \otimes U_x + \ketbra{x}{x+1}\otimes U_x^{-1}.
\end{align}
Each term of the Hamiltonian involves two nearest neighbor sites of the clock and a unitary operator $U_x$ acting on the register.  The ordered product $U_{s-1} \ldots U_2 U_1$ realizes some input/output transformation we want the quantum computing device to accomplish. \Fref{fig:system} shows the architecture of the machine.
\\Because of the properties of the Hamiltonian $H_F$, the position of the excitation along the chain uniquely determines the state of the register. This fact has interesting consequences. Consider an initial condition of the form $\ket{\psi_0}=\ket{1}\otimes\ket{R_1}\equiv \ket{1,R(1)}$, i.e. with the particle located at the beginning of the chain and the register in an input state $\ket{R_1}$. Then the set
\[
\mathcal{B}(\psi_0)=\{ \ket{1,R_1}, \ket{2,R_2},\ldots,\ket{s,R_s}\},
\]
where $\ket{R_j} = U_{j-1} \ldots U_2 U_1 \ket{R_1}$, constitutes an orthonormal basis, the \emph{computational basis}, or \emph{Peres basis} \cite{peres85}, for the region of  Hilbert space  visited by the evolved state $\ket{\psi_t} = \exp(-i H_F t)\ket{\psi_0}$. We refer to the space spanned by the Peres basis as to the \emph{computational subspace}. 
\\In particular, if upon measurement the cursor is found at the rightmost site of the chain, the register collapses to the output state $\ket{R(s)} = U_{s-1} \ldots U_2 U_1 \ket{R(1)}$. 
\\[5pt]It is clear that the capability of the chain of transferring the excitation from one end to the other is a central matter in the analysis of the computational power of the Feynman machine.
\\Before discussing the kinematics of the cursor, we point out that the sole effect of the interaction of the clock with the register (of $n$ spins) is the appearance of a degeneracy of order $2^n$ in the spectrum $\{e_k\}_{k=1}^s$ of the tight-binding (clock) Hamiltonian $H_0$ (see \eref{eq:H0}). Once an initial condition of the form $\ket{\psi_0}=\ket{1,R(1)}$ has been set, however, the spectrum of the Hamiltonian $H_F$ restricted to the computational subspace is no longer degenerate. In this subspace, to each eigenvalue $e_k$ (see \eref{eq:freeEval}) there corresponds the eigenvector
 \begin{align}
 &\ket{v_k} = \sqrt{\frac{2}{s+1}}\sum_{x=1}^s \sin\left (\frac{k \pi x}{s+1} \right) \ket{x,R(x)}.
\end{align}
The presence of a (random) potential acting on the cursor can drastically reduce the probability of finding the excitation at the end of the chain (i.e. the computation completed) but does not alter the working mechanism of the Feynman machine. 
\\For the sake of definiteness, let  the state \ket{\psi_0} evolve under the Hamiltonian
\begin{align}\label{eq:FeynPot}
 H_{F,V}= H_F + V = H_F+\sum_{x=1}^s f(x) \ketbra{x}{x}
\end{align}
where $V$ represents a potential. The projector on the space $\mathcal{H}_{\psi_0}$ spanned by the Peres basis $\mathcal{B}(\psi_0)$
\[
\mathcal{P}(\mathcal{H}_{\psi_0})=\sum_{x=1}^s \ketbra{x,R_x}{x,R_x}
\]
remains a constant of the motion. Indeed
\begin{align} \label{eq:conservation}
& \left [V,\mathcal{P}(\mathcal{H}_{\psi_0}) \right] =\left [ \sum_x f(x)\ketbra{x}{x},\mathcal{P}\llrr{\mathcal{H}_{\psi_0}} \right ] = \\
&=\sum_{x,y} f(x)\left[\ \ketbra{x}{x},\ketbra{y,R_y}{y,R_y} \ \right] = \nonumber \\
&=\sum_{x,y}f(x) \left(\ket{x}\braket{x}{y,R_y} \bra{y,R_y} \right.\nonumber \\
& \left. -\ket{y,R_y}\braket{y,R_y}{x}\bra{x}=0 \right). \nonumber
\end{align}
The study of the evolution of the Feynman machine can therefore be reduced to the analysis carried out in section \ref{sec:model}. In particular, the Hamiltonian \eref{eq:FeynPot}, restricted to the subspace $\mathcal{H}_{\psi_0}$, and represented in the same basis, admits $s$ eigenvalues $e_k$ and corresponding eigenvectors $\ket{e_k}$.
%
\\[5pt]By requiring the operators $U_x$ to act on a single register qubit, the Hamiltonian \eref{eq:feyn} becomes 3-local. This restriction simplifies the implementation of the device; but at the same time it reduces the computational capability of the Feynman machine: it does not allow the realization of any two-qubit gates.  In order to restore the universality of the model Feynman introduced the \emph{Switch} circuit \cite{feyn86}, that implements the selection statement. For a recent review, see \cite{nagaj10}. 
\\[5pt]In what follows, we consider a particular instance of the Switch:  the \emph{Controlled NOT} (CNOT) gate. The gate acts on two register qubits: $\sigma(c)$, the \emph{controlling} qubit and $\sigma(p)$, the \emph{controlled} (or \emph{passive}) qubit. In our implementation, if the controlling qubit is ``up'', namely $\sigma_3(c)=+1$, the passive qubit $\sigma(p)$ is ``negated'' by applying $\sigma_1(p)$; if the controlling qubit is ``down'' ($\sigma_3(c)=-1$) then the controlled qubit is left unchanged.
\\The Hamiltonian realizing the gate is:
\begin{widetext}
\begin{align}
H_{CNOT}(a)=-\frac{1}{2}\llrr{\ketbra{a+1}{a}\otimes \frac{1+\sigma_3(c)}{2} + \ketbra{a+2}{a+1}\otimes \sigma_1(p) + \ketbra{a+5}{a+2}\otimes \frac{1+\sigma_3(c)}{2}}  \\
-\frac{1}{2}\llrr{\ketbra{a+3}{a}\otimes \frac{1-\sigma_3(c)}{2} + \ketbra{a+4}{a+3} + \ketbra{a+5}{a+4}\otimes \frac{1-\sigma_3(c)}{2}}+ \mbox{H.c.}\nonumber
\end{align}
\end{widetext}
Here $a$ indicates the site of the clock at which the switch starts and H.c. stands for Hermitian conjugate. 
\\The state of the controlling qubit determines which branch of the CNOT circuit is visited: if $\sigma(c)$ is ``up'', the excitation will walk along the upper path and $\sigma_1(p)$ is applied. If $\sigma(c)$ is ``down'', the excitation will walk along the lower path and no operator is applied to the register.
\\In order to discuss the effects of imperfections and dissipation on such a circuit, it is expedient of put some ``inertial'' sites before and after the CNOT, ``during'' which we do not apply any transformation to the register qubits. The full Hamiltonian of the circuit is:
\begin{align} \label{eq:CNOTtel}
 H_C&=\sum_{x=1}^{a-1} -\frac{1}{2} \llrr{\ketbra{x+1}{x} + \ketbra{x+1}{x}} \\
 & +H_{CNOT}(a)  \nonumber \\
 &-\frac{1}{2} \sum_{x=a+5}^{s-1}\llrr{\ketbra{x+1}{x} + \ketbra{x+1}{x}} \nonumber
\end{align}
\Fref{fig:CNOT} shows the structure of the CNOT gate. 
\\If the initial state of the system is 
\begin{align}
\ket{\psi_0^U}&=\ket{Q=1}\otimes \ket{\sigma_3(c)=+1, \sigma_3(p)=-1}= \nonumber\\
&=\ket{1} \otimes \ket{+1,-1}. \nonumber
\end{align}
 only the upper branch of the CNOT is visited during the evolution of the system and, if the cursor reaches the region $a+5,\ldots,s$ the state of the  register is \ket{+1,+1}. The analysis of the motion of the cursor is once more simplified by the use of the Peres basis
\begin{align}
\mathcal{B}(\psi_0^U)= &\left ( \ket{1}\otimes \ket{+1,-1},\ldots,\right. \nonumber \\
&\ket{a+1}\otimes \ket{+1,-1}, \ket{a+2}\otimes \ket{+1,+1}, \nonumber \\
&\left. \ket{b}\otimes \ket{+1,+1},\ldots,\ket{s}\otimes \ket{+1,+1}\right). \nonumber
\end{align}
We indicated by  $b=a+5$ the position at which the CNOT ends.
\\A similar argument applies to the initial condition $\ket{\psi_0^D}=\ket{1}\otimes\ket{-1,-1}$. With this initial condition it is the lower branch of the CNOT that is visited. The corresponding computational subspace is spanned by 
\begin{align}
\mathcal{B}(\psi_0^D)= &\left ( \ket{1}\otimes \ket{1,-1},\ldots, \right. \nonumber \\
& \ket{a+3}\otimes \ket{-1,-1},\ket{a+4}\otimes \ket{-1,-1}, \nonumber \\
&\left. \ket{b}\otimes \ket{-1,-1},\ldots,\ket{s}\otimes \ket{-1,-1}\right). \nonumber
\end{align}
Both bases consist of $s-2$ orthogonal elements. The Hamiltonian \eref{eq:CNOTtel} restricted to $\mathcal{H}_{\psi_0^U}$ (or $\mathcal{H}_{\psi_0^D}$) and represented in the basis $\mathcal{B}\llrr{\psi_0^U}$ (or $\mathcal{B}\llrr{\psi_0^D}$)  is equivalent to the ``free'' Hamiltonian \eref{eq:H0} on $s-2$ sites. In particular, the motion of the cursor along the two computational paths is exactly the same.
%
The presence of random imperfections, namely of independent realization $\epsilon_1,\epsilon_2,\ldots, \epsilon_{s}$ of a zero-mean Gaussian random variable $R$ with variance $\sigma^2$, perturbs the motion along the computational paths. Since $\epsilon_{a+1} \neq \epsilon_{a+3}$ and $\epsilon_{a+2} \neq \epsilon_{a+4}$ with probability 1, the spectra of the Hamiltonians restricted to $\mathcal{H}_{\psi_0^U}$ and $\mathcal{H}_{\psi_0^D}$ will be slightly different. But the correctness of the computation is guaranteed by the conservation law \eref{eq:conservation}.
\\The problem is the ensuing localization that we discussed in section II; it can suppress the probability of reaching the region $b,\ldots,s$. 
\\[5pt]In the previous section we showed that we can make the excitation  travel beyond the localization length, and actually reach the rightmost sites of a linear lattice, by means of an external force and dissipation. In what follows we show that this result extends straightaway to \emph{each} computational path.
\\[5pt]We add an external field such that the potential difference between one site and its right next-neighbor(s) is $-g$, by setting
\begin{align}
 V_L'= -\llrr{\sum_{x=1}^{a+2} g x \ketbra{x}{x} + \sum_{x=a+3}^s  g (x-2) \ketbra{x}{x}}. \nonumber
\end{align}
The Hamiltonian evolution of the system is given by
\begin{align} \label{eq:hamAndPotCNOT}
	&H_{CNOT} + V_R + V_L'=  H_{CNOT} + \sum_{x=1}^s \epsilon_x \ketbra{x}{x} \\
	& -\llrr{\sum_{x=1}^{a-2} g x \ketbra{x}{x} + \sum_{x=a+3}^s  g (x-2) \ketbra{x}{x}}. \nonumber
\end{align}
The initial condition of the system (either \ket{\psi_0^U} or \ket{\psi_0^D}) selects the relevant computational subspace (either $\mathcal{H}_{\psi_0^U}$ or $\mathcal{H}_{\psi_0^D}$). By switching on an interaction  with a thermal bath satisfying the hypotheses listed in \sref{sec:noise-assisted}, we determine a walk on the energy landscape of the selected space identical to the one we described in the previous section. For example, if the initial state of the machine is $\rho_0^U=\ketbra{\psi_0^U}{\psi_0^U}$, the elements of the density matrix $\rho^U(t)$, restricted to the computational subspace $\mathcal{H}_{\psi_0^U}$ and  there expressed in the energy representation, are given by \eref{eq:off} and \eref{eq:trans}.
\\This means that the clocking particle can travel beyond the CNOT. And when it does, the conservation law \eref{eq:conservation} guarantees the state of the register is the correct one. 
\\We can thus state that noise can assist \emph{classical} computation on a \emph{quantum} computing device.
\\\Fref{fig:figNAC} shows the probability of reaching, starting from site 1, the region $b,\ldots,s$ under different evolutions. The presence of imperfections along the chain strongly reduces the probability of ever reaching the sites located \emph{after} the gate. For the ballistic evolution there are times at which this probability is close to unity. When the evolution is dissipative, the time required to pass the CNOT is one order of magnitude larger than the first ``arrival time'' under Hamiltonian evolution. But the probability of finding the computation completed is monotonically increasing to one.
\\[5pt]The  equivalence between successful transport and successful computation does not extend to the case that sees the controlling qubit $\sigma(c)$ in a superposition of ``up'' ($\sigma_3(c)=+1$) and ``down'' ($\sigma_3(c)=-1$).
\\Suppose, for instance, that the controlling spin $\sigma(c)$ starts from the state 
\[
\ket{\sigma_1(c)=+1} = \frac{\ket{\sigma_3(c)=+1}+ \ket{\sigma_3(c)=-1}}{\sqrt{2}},
\]
and therefore the complete system starts from the initial condition
\[
\ket{\psi_0^{U+D}}=\frac{\ket{\psi_0^U}+\ket{\psi_0^D}}{\sqrt{2}}.
\]
If the evolution were purely Hamiltonian, with identical on-site energies in the two branches of \fref{fig:CNOT}, application of the CNOT primitive would bring the register \llrr{\sigma(c),\sigma(p)} to the maximally entangled Bell state \cite{nielsen}
 \[
 \ket{\Phi^+}=\frac{\ket{-1,-1}+\ket{+1,+1}}{\sqrt{2}}.
 \]
The presence of random imperfections on the circuit we are considering produces two effects:
\begin{itemize}
\item the introduction of random relative phases between the two computational paths. Because of the relative phases, when the particle is on the right of the CNOT the state of the register can be slightly different from the target Bell state. But this effect can in principle be ``bounded away'' for small values of the variance $\sigma^2$;
\item Anderson localization of the cursor; this is the key issue that we discuss below.
 \end{itemize}
%
%
%
The transmission of the excitation from one end to the other of the circuit can be assisted by dissipation. But this time dissipation destroys the computation. The density matrix of the system can be schematically represented as a block matrix:
\begin{align}
\left( 
\begin{array}{c|c}
\rho_{DD}(t) & \rho_{DU}(t)  \\
\hline
\rho_{UD}(t) & \rho_{UU}(t)
\end{array} 
\right).
\end{align}
The blocks $\rho_{DD}(t)$ and $\rho_{UU}(t)$ represent the evolution of the projections of the initial state respectively on $\mathcal{H}_{\psi_0^D}$ and $\mathcal{H}_{\psi_0^U}$. The blocks $\rho_{DU}(t)$ and $\rho_{UD}(t)$ correspond to the coherences \emph{between the computational subspaces}.
Lindblad evolution, according to \eref{eq:off}, destroys all the coherences. This does not affect the computation within each computational subspace: the decoherence simply transforms a quantum walk on the energy landscape into a random walk. But the damping of coherences between the computational subspaces destroys the desired entanglement; it transforms the state of the register into the maximally mixed state
\[
\frac{\ketbra{-1,-1}{-1,-1}+\ketbra{+1,+1}{+1,+1}}{2}.
\]
This is evident from the fact that the von Neumann entropy of the register for a machine evolving from \ket{\psi_0^{U+D}} tends to $\log(2)$ as $t \to \infty$ (solid line in \fref{fig:figFidelityEntropy}). For comparison, the dashed line of \fref{fig:figFidelityEntropy} shows that, starting from \ket{\psi_0^U}, the entropy of the register tends to the value 0 competing to a pure state.
\\[5pt] For both initial conditions the entropy reaches a maximum about the time $\bar{t}$ when $E(Q)$ is close to $s/2$; at this time $\sqrt{var(Q)}$ is large (see \fref{fig:figDiss}(b)).
\\For the initial condition $\ket{\psi_0^U}$, the state of the register at time $\bar{t}$ is a mixture of two states: \ketbra{+1,-1}{+1,-1} (cursor on the upper branch and at the left of the NOT) and \ketbra{+1,+1}{+1,+1} (cursor on the upper branch and at the right of the NOT). Hence the value $\log(2)$ for the absolute maximum at time $\bar{t}$ of the entropy for the dashed line of \fref{fig:figFidelityEntropy}.
\\If the controlling qubit $\sigma(c)$ is in a superposition of ``up'' and ``down'', at time $\bar{t}$ the register is in a mixture of three states: \ketbra{+1,-1}{+1,-1} (cursor on the upper branch and at the left of the NOT) with weight 1/4, \ketbra{+1,+1}{+1,+1} (cursor on the upper branch and at the right of the NOT) with weight 1/4 \emph{and}, this time, \ketbra{-1,-1}{-1,-1} (cursor on the lower branch) with weight 1/2. Hence the value $(3/2) \log(2)$ of the maximum reached by the solid line.
\\[5pt]For the initial condition \ket{\psi_0^{U+D}}  the steep growth of the entropy from 0 to $\log(2)$ that leads to the plateau around time $t=40$  is a witness of the disruption of coherences between computational subspaces, as it is easy to check by direct inspection of the decay rate of the explicit solution of Eqn.\ref{eq:off} for the coherences.


\section{Conclusions and outlook} \label{sec:conclusions}

We presented an example of noise-assisted transport on a lattice and discussed a set of hypotheses that allow for the extension of noise-assisted transport to noise assisted computation. We showed that it is indeed possible to exploit the interaction of the (Feynman) computing device with a reservoir to assist \emph{classical} computation on a quantum device. On the other side, the suppression of coherences due to the presence of a bath destroys entanglement.
\\[5pt]We are presently exploring the possibility, offered by Time-Dependent Density Functional Theory, of simulating the position probability distribution of an open  quantum systems with the one of a system undergoing unitary propagation under a time dependent potential \cite{tempel10}. The interest of this proposal is in the fact (made possible by the extension of the point of view of TDDFT to discrete systems \cite{aspuru11,tokatly12} and to open discrete systems \cite{defa12}) that the same external time-dependent potential may be made to act along two branches of the computation, leading to completion without loss of phase information. 

\end{document}